# Optomechanically-induced spontaneous symmetry breaking

Mohammad-Ali Miri[1], Ewold Verhagen[2], and Andrea Alù[1]

[1]*Department of Electrical and Computer Engineering, The University of Texas at Austin, Austin, TX 78712, USA*

[2]*Center for Nanophotonics, AMOLF, Science Park 104, 1098 XG Amsterdam, The Netherlands*

**Abstract**

*We explore the dynamics of spontaneous breakdown of mirror symmetry in a pair of identical optomechanical cavities symmetrically coupled to a waveguide. Large optical intensities enable optomechanically-induced nonlinear detuning of the optical resonators, resulting in a pitchfork bifurcation. We investigate the stability of this regime and explore the possibility of inducing multistability. By injecting proper trigger pulses, the proposed structure can toggle between two asymmetric stable states, thus serving as a low-noise nanophotonic all-optical switch or memory element.*

Symmetry is a tantalizing concept in modern physics, governing many of its fundamental laws [1]. Beyond its crucial role in the context of theoretical physics, symmetry is important in several areas of applied physics, including photonics, as symmetry and its breaking can be fruitfully utilized to design photonic structures with desired properties. The symmetry groups of the eigenfunctions in photonic crystals, for example, directly affect their optical responses [2]. Such spatial symmetries thus have been exploited to design optical cavities and channel drop filters [3]-[4]. Symmetry and symmetry-breaking principles have also been explored in chiral metamaterials [5] and in designing micro- and nano-lasers [6]-[7]. In general, symmetry breaking can occur in explicit or spontaneous forms. In the latter scenario, an initially symmetric state evolves into an asymmetric one even though the governing dynamical equations remain invariant under symmetry transformations. Spontaneous breaking of symmetry has proven a particularly powerful concept with wide implications in physics, ranging from the Higgs mechanism to Josephson junctions [8]-[9].

One of the simplest and most explored examples of symmetry in quantum mechanics is the spatial mirror symmetry associated with two identical and closely spaced quantum wells. Due to the underlying parity, the eigenstates associated with such a system are symmetrically distributed around the center of the two wells. In the presence of nonlinearities, however, the situation can be made very different. In this case, as a result of pitchfork bifurcation arising at high enough intensities, the system undergoes spontaneous symmetry breaking, and the wavefunctions are no longer evenly distributed [10]. This concept is not limited to quantum mechanics and has been investigated theoretically and experimentally in a range of nonlinear optical systems, such as Fabry-Perot resonators, coupled waveguides, and photonic crystal defect cavities [7],[11]-[20]. Despite different structures and geometries, the symmetry-breaking phenomena reported so far have been all based on utilizing intrinsic material nonlinear responses.

Here, we explore how spontaneous mirror symmetry breaking between two optical modes can be initiated by the back-action of optical radiation on the mechanical degrees of freedom. Spurred by advances in fabrication of high-quality optical and mechanical resonators, cavity optomechanics has recently attracted considerable attention [21]-[22]. Cavity-optomechanical systems enable exploiting strong interactions between optical fields and mechanical vibrations mediated through radiation pressure. The mutual interaction between light and motion has led to the observation of phenomena affecting light propagation such as electromagnetically-induced transparency and slow light [23]-[24], as well as mechanisms to control mechanical motion such as dynamical back-action cooling and parametric amplification [25]-[28]. It has been long known that optomechanical coupling can mimic an effective Kerr-type nonlinearity [29], which can result in classical and quantum nonlinear phenomena such as optical bistability [30] and sub-Poissonian light [31]. Such strong and concentrated nonlinear effects, that can exceed even thermal nonlinearities in strength [32]-[33], paired with a low-noise platform, opens useful applications for light manipulation in nanophotonic devices [34].

In the following, we explore how the optomechanical nonlinearity can serve to induce the spontaneous breakdown of the mirror symmetry between two identical coupled optical cavities that are symmetrically excited via a bus waveguide. Importantly, the triggering of symmetry breaking by optomechanical interactions can lead to rich physical responses, due to their highly resonant and dynamic nature of this multi-physics coupling. In the following, we show that optical frequency detuning and losses play an important role in symmetry breaking, and we analytically find the conditions under which symmetry may be broken and optimally induced in these systems. In addition, we show how the proposed structure can support multistability for certain parameter ranges. The stability of the steady-state solutions is investigated, showing that the proposed structure can exhibit bistability between a degenerate pair of asymmetric states in a regime where the symmetric eigenstate is unstable. Finally, the associated dynamics of the proposed structure is explored and potential applications for low-noise nanophotonic switching and memory are discussed.

Figure 1 schematically shows an arrangement of two identical optomechanical resonators symmetrically coupled through a bus waveguide. Although this figure shows microring resonators, the following formulation is quite general, and it can be applied to other types of optomechanical systems. The dynamics of this system is described through the coupled mode equations

$$\frac{da_1}{dt} = \left(i(\Delta + Gx_1) - \frac{\kappa}{2}\right)a_1 - \frac{\kappa_e}{2}a_2 + \sqrt{\kappa_e}s_{\text{in}}, \quad (1.a)$$

$$\frac{d^2x_1}{dt^2} = -\Omega_m^2 x_1 - \Gamma_m \frac{dx_1}{dt} + \frac{\hbar G}{m}|a_1|^2, \quad (1.b)$$

$$\frac{da_2}{dt} = \left(i(\Delta + Gx_2) - \frac{\kappa}{2}\right)a_2 - \frac{\kappa_e}{2}a_1 + \sqrt{\kappa_e}s_{\text{in}}, \quad (1.c)$$

$$\frac{d^2x_2}{dt^2} = -\Omega_m^2 x_2 - \Gamma_m \frac{dx_2}{dt} + \frac{\hbar G}{m}|a_2|^2, \quad (1.d)$$

where, in these relations, $a_{1,2}$ represent the amplitude of the two optical modes, such that $|a_{1,2}|^2$ is the total number of photons inside each cavity, $\Delta = \omega_L - \omega_0$ shows the detuning of the driving laser frequency $\omega_L$ with respect to the resonance frequency of the optical cavity $\omega_0$, $G$ is the optical frequency shift per unit of displacement and $s_{\text{in}}$ is a measure of the amplitude of the driving laser such that $|s_{\text{in}}|^2$ is the incoming photon flux. The decay rate $\kappa = \kappa_\ell + \kappa_e$ represents the total optical losses, including absorption and scattering $\kappa_\ell$ and the out-coupling loss $\kappa_e$. On the mechanical side, $x_{1,2}$ shows the mechanical displacements of the two cavities, $\Omega_m, \Gamma_m$ and $m$ represent the resonance frequency, decay rate and mass of the mechanical modes, and finally $\hbar$ is the Planck constant.

Under steady state conditions, we consider fixed point solutions $(\bar{a}_1, \bar{x}_1)$ and $(\bar{a}_2, \bar{x}_2)$ for the two optomechanical systems, where $\bar{a}_{1,2}$ represent the steady state solution of the optical modes $a_{1,2}$ inside the two cavities and $\bar{x}_{1,2}$ represent their associated mechanical displacements. By ignoring all time derivatives in Eqs. (1), we find

$$\left(i(\Delta + \gamma|\bar{a}_1|^2) - \frac{\kappa}{2}\right)\bar{a}_1 - \frac{\kappa_e}{2}\bar{a}_2 + \sqrt{\kappa_e}s_{\text{in}} = 0, \quad (2.\text{a})$$

$$\left(i(\Delta + \gamma|\bar{a}_2|^2) - \frac{\kappa}{2}\right)\bar{a}_2 - \frac{\kappa_e}{2}\bar{a}_1 + \sqrt{\kappa_e}s_{\text{in}} = 0, \quad (2.\text{b})$$

where, $\gamma = \frac{\hbar G^2}{m\Omega_m^2}$ is the optomechanically-induced steady state cubic nonlinearity coefficient, and we have $\bar{x}_1 = \frac{\hbar G}{m\Omega_m^2}|\bar{a}_1|^2$ and $\bar{x}_2 = \frac{\hbar G}{m\Omega_m^2}|\bar{a}_2|^2$. After setting these two relations equal, and by defining the intensities $A_1 = |\bar{a}_1|^2$ and $A_2 = |\bar{a}_2|^2$, we can show that

$$(\gamma^2(A_1^2 + A_1 A_2 + A_2^2) + 2\gamma\Delta(A_1 + A_2) + \Delta^2 + \kappa_\ell^2/4)(A_1 - A_2) = 0. \quad (3)$$

As expected, for all sets of parameters this equation admits a symmetric solution $A_1 = A_2$. However, for some range of parameters asymmetric solutions $A_1 \neq A_2$ also arise. Inspecting Eq. (3), and considering that $A_1$ and $A_2$ are both positive quantities, we find that asymmetric solutions require $\Delta < 0$, i.e., operating in the red-detuned regime, which ensures the absence of parametric instabilities of the mechanical oscillator as long as the input power does not exceed a critical level [35]. To find the exact parameter range required for asymmetric solutions, we solve Eq. (3) for $A_2$, which results in $2\gamma A_2 = -\gamma A_1 -$ $2\Delta \pm \sqrt{D}$ where $D = -\kappa_\ell^2 - 4\Delta\gamma A_1 - 3\gamma^2 A_1^2$. To have valid solutions, $D$ should be positive, which happens only for

$$\Delta < -\frac{\sqrt{3}}{2}\kappa_\ell. \quad (4)$$

Although this necessary condition for symmetry breaking depends only on the frequency detuning and intrinsic optical losses, it is expected to depend also on the input power level. In fact, when condition (4) is satisfied, the symmetry breaking threshold of intra-cavity photon numbers can be obtained by solving the asymmetric branch of Eq. (3) for $A_1 = A_2$, which results in

$$A_{\text{th}}^\pm = \frac{1}{6\gamma}\left(-4\Delta \mp \sqrt{4\Delta^2 - 3\kappa_\ell^2}\right), \quad (5)$$

associated with the lower ($-$) and upper ($+$) bifurcation points of the bistability region. The critical input power level at which symmetry breaking begins and ends can be obtained by solving Eq. (2) for $\bar{a}_1 = \bar{a}_2$ and using the threshold intra-cavity photon numbers obtained from Eq. (5). This leads to the threshold input photon flux:

$$\left|s_{\text{th}}^\pm\right|^2 = \frac{1}{\kappa_e}\left[\left(\Delta + \gamma A_{\text{th}}^\pm\right)^2 + \left(\frac{\kappa+\kappa_e}{2}\right)^2\right]A_{\text{th}}^\pm. \quad (6)$$

Figure 2 shows the steady state solutions of Eqs. (2) as a function of the input photon flux for different frequency detunings. For this example, we have considered silica microtoroid resonators supporting a mechanical radial breathing mode, evanescently coupled to a tapered fiber [36]. Here, we assume $\kappa/2\pi = 2\kappa_e/2\pi = 2\kappa_\ell/2\pi = 1$ MHz, $\Omega_m/2\pi = 50$ MHz, $\Gamma_m/2\pi = 10$ KHz, $G/2\pi = 6$ GHz/nm, $m = 6$ ng. Such parameters are within experimental reach (see for example [37],[38]). As shown in Fig. 2(a), for the case $\Delta = 0$ the only possible solution is the symmetric eigenstate. By decreasing the detuning parameter below the critical point $\Delta_{th} = -\sqrt{3}\kappa_\ell/2$, a bifurcation emerges for sufficiently large input powers. This is shown in Figs. 2(b) for $\Delta = -\kappa_\ell$, where the asymmetric solutions appear between two bifurcation points associated with the critical input photon flux levels $|s_{in}^-|^2 \approx 0.49 \times 10^{14}$ and $|s_{in}^+|^2 \approx 0.74 \times 10^{14}$ s$^{-1}$. For a detuning rate $\Delta = -1.5\kappa_\ell$ (Fig 2(c)), the bifurcation pattern changes, as each branch of the asymmetric solutions involves unstable branches. By further decreasing the detuning to $\Delta = -4\kappa_\ell$ (Fig 2(d)), the bifurcation pattern becomes even more complex, since optical bistability becomes the dominant effect. As shown in the following section, in this case both the unbroken and broken symmetry states are stable, while in the asymmetric mode, a large contrast between photon numbers in the two cavities can be achieved.

The stability of the derived fixed point solutions can be investigated by evaluating the eigenvalues of the associated Jacobian matrix. Defining the normalized momenta $p_{1,2} = dx_{1,2}/dt$, we first reduce the mechanical equation of motion to first-order equations. After defining perturbed solutions $a_{1,2} = \bar{a}_{1,2} + \delta a_{1,2}$, $x_{1,2} = \bar{x}_{1,2} + \delta x_{1,2}$ and $p_{1,2} = \bar{p}_{1,2} + \delta p_{1,2}$, the equations of motion can be linearized around the fixed point solutions $\bar{a}_{1,2}, \bar{x}_{1,2}, \bar{p}_{1,2}$ where $\bar{x}_{1,2} = \frac{\hbar G}{m\Omega_m^2}|\bar{a}_{1,2}|^2$ and $\bar{p}_{1,2} = 0$. The evolution equations of the perturbed scenario can be written as

$$\frac{d}{dt}\begin{pmatrix}\delta\psi_1\\ \delta\psi_2\end{pmatrix} = \begin{pmatrix}\mathcal{L}_1 & \mathcal{M}\\ \mathcal{M} & \mathcal{L}_2\end{pmatrix}\begin{pmatrix}\delta\psi_1\\ \delta\psi_2\end{pmatrix}, \quad (7)$$

where $\delta\psi_{1,2} = (\delta a_{1,2}, \delta a_{1,2}^*, \delta x_{1,2}, \delta p_{1,2})^t$, and the blocks of the Jacobian matrix are defined as

$$\mathcal{L}_{1,2} = \begin{pmatrix} i\bar{\Delta}_{1,2} - \kappa/2 & 0 & +iG_{1,2} & 0 \\ 0 & -i\bar{\Delta}_{1,2} - \kappa/2 & -iG_{1,2}^* & 0 \\ 0 & 0 & 0 & 1/m \\ \hbar G_{1,2}^* & \hbar G_{1,2} & -m\Omega_m^2 & -\Gamma_m \end{pmatrix}, \quad (8.a)$$

$$\mathcal{M} = -\frac{\kappa_e}{2}\begin{pmatrix} 1 & 0 & 0 & 0 \\ 0 & 1 & 0 & 0 \\ 0 & 0 & 0 & 0 \\ 0 & 0 & 0 & 0 \end{pmatrix}, \quad (8.b)$$

where $\bar{\Delta}_{1,2} = \Delta + G\bar{x}_{1,2}$ and $G_{1,2} = G\bar{a}_{1,2}$ respectively represent the modified detuning and the enhanced optomechanical frequency shifts of the two cavities. As a result of the dynamical perturbation equations (7), a fixed point solution is stable as long as all eigenvalues of the associated Jacobian matrix exhibit a negative real part. This condition can be numerically investigated for all steady-state solutions of Fig. 2. Figure 3 shows all eight eigenvalues of the Jacobian for the symmetric and asymmetric solutions of Fig. 2(b). The only portion with unstable eigenvalues, shown with a dashed line, corresponds to the symmetric eigenstates in the region where the asymmetric eigenstates exist. The stable and unstable regions for different cases in Fig. 2 are shown with solid and dashed lines, respectively. Interestingly, for a certain parameter range and at specific input power levels, the proposed structure exhibits multistability. As shown in Fig. 3, for $1.6 \times 10^{14} \lesssim |s_{in}|^2 \lesssim 3.4 \times 10^{14}$ and $5 \times 10^{14} \lesssim |s_{in}|^2 \lesssim 5.3 \times 10^{14}$, we find three stable solutions, and for $3.4 \times 10^{14}$ s$^{-1} \lesssim |s_{in}|^2 \lesssim 5 \times 10^{14}$ s$^{-1}$ four stable eigenstates coexist.

The stability of the fixed point solutions can be further explored dynamically by directly simulating the evolution (1), as shown in Fig. 4. Here, the results are presented for the optomechanical system of Fig. 2(b) at two different photon flux levels $|s_{in}|^2 = 0.6 \times 10^{14}$ (Fig. 2(a,b)) and $|s_{in}|^2 = 0.4 \times 10^{14}$ s$^{-1}$ (Fig. 2(c,d)), which correspond to stable asymmetric and symmetric regimes respectively. In both cases, the fixed point solutions of Eqs. (2) are attractors for arbitrary initial excitations of the two cavities.

The proposed structure can operate as an all-optical memory element, switching between its two stable asymmetric states when triggered by weak control pulses to one of the two cavities such that the state of the system can hop to the basin of attraction of the other stable state. In order to toggle between the two states, when applied to the lower intensity cavity the pulse should be positive and when applied to the higher intensity cavity it should be negative. Alternatively, one can apply a positive pulse control to either cavity that is in its lower intensity state. Figure 5 shows time-domain simulations for design parameters similar to those used in Fig. 2(c), while both cavities are initially populated with a stable state. Interestingly, the intensity contrast between the two switching states can be easily controlled via the frequency detuning $\Delta$. This can be shown by solving the asymmetric branch of Eq. (3) for a fixed point solution that results in the maximum contrast $|A_2 - A_1|$. By enforcing the condition $d(|A_2 - A_1|)/dA_1 = 0$, the maximum contrast is

$$\max(|A_2 - A_1|) = \frac{1}{\sqrt{3}\gamma}\sqrt{4\Delta^2 - 3\kappa_\ell^2}, \quad (9)$$

which is a monotonically increasing function of the frequency detuning. An upper limit on the toggling time between these two stable states can be approximated by $t_0 = 1/\Re(\lambda_0)$ where $\lambda_0$ represents the Jacobian eigenvalue with the algebraically largest real part. Even though for the example presented in Fig. 5 this limit is $t_0 \approx 36\mu m$, in principle the switching occurs in a few microseconds.

In the analysis presented so far, the two optomechanical cavities are assumed to be identical while in practice imperfections may arise in various parameters ($\Delta$, $\gamma$, $\kappa_\ell$, and $\kappa_e$), thus breaking the parity inversion symmetry of the steady state equations (2). In order to investigate the effect of such imperfections, we break the mirror symmetry of the problem by considering two different optomechanical nonlinearity rates $\gamma_{1,2} = (1 \pm \varepsilon)\gamma$, with $\varepsilon \ll 1$, for the two cavities and obtain the nonlinear fixed points as shown in Fig. 6. As expected, a small perturbation ($\varepsilon = 0.002$) lifts the degeneracy of the nonlinear eigenstates leading into a coexisting pair of asymmetric eigenstates with slightly different on/off intensities in the two cavities (Fig. 6(a)). Similar to the previous case, this system supports an unstable eigenstate with minor intensity contrast between the two cavities due to the lifted degeneracy. As shown in Fig. 6(b), by increasing the detuning $\varepsilon$, the bistability region shrinks and eventually evaporates above a critical choice of $\varepsilon$ (see Fig. 6(c)). In the latter scenario, the symmetry-breaking signature appears as a large intensity contrast between the two cavities for a specific power range.

It is worth stressing that the effective static nonlinearity offered in an optomechanical system can exceed that of Kerr-type nonlinear resonators, which tend to suffer from large intrinsic losses as nonlinear effects grow [39]. The optomechanically-induced frequency shift per photon $\gamma = \partial\omega/\partial\bar{n} = \hbar G^2/m\Omega_m^2$ can be rewritten in terms of the single photon optomechanical coupling rate $g_0 = G x_{\text{ZPF}}$, with $x_{\text{ZPF}} = \sqrt{\hbar/2m\Omega_m}$ representing the mechanical zero point fluctuation amplitude, as $\gamma = 2 g_0^2/\Omega_m$. The quantity $g_0^2/\Omega_m$ represents the strength of the mechanically-assisted photon-photon interaction, which can be significantly large in suitably designed optomechanical systems [22], thus supporting strong nonlinear frequency detunings at low intensities. For example, using a nanophotonic photonic-crystal-based implementation with the parameters presented in ref. [40] would yield a symmetry-breaking threshold at only 830 intracavity photons, for cavity linewidths of 0.5 THz. Such large linewidths would facilitate straightforward frequency matching of the two cavities. In this regard, optomechanical cavities offer an exciting route for inherently low-power and low-noise nonlinear nanophotonic switching devices and memories. We are currently exploring the impact of thermomechanical noise on the operation of these devices.

Finally, it should be noted that the optical bistability achieved under spontaneous symmetry breaking in the proposed coupled cavity structure occurs at lower power levels compared to the bistability behavior in a single optomechanical cavity. In fact, for an optomechanical cavity described in steady state with $(i(\Delta + \gamma|\bar{a}|^2) - \kappa/2)\bar{a} + \sqrt{\kappa_e}s_{\text{in}} = 0$, the necessary condition for bistability is found to be $\Delta < -\frac{\sqrt{3}}{2}\kappa$, while the two bistability turning points $A_{\text{th}}^{\pm}$, associated with $d|s_{\text{in}}|^2/d|\bar{a}|^2 = 0$, are found to be $A_{\text{th}}^{\pm} = \frac{1}{6\gamma}(-4\Delta \mp \sqrt{4\Delta^2 - 3\kappa^2})$. Clearly, in this case, larger frequency detunings are required to reach bistability which in turns requires larger intraccavity photon numbers.

To conclude, we have shown that a coupled arrangement of identical optomechanical cavities can undergo spontaneous symmetry breaking in the red detuning regime for low input power levels, which may be triggered and controlled by suitable input pulses. We studied the static and dynamic behavior of this system and explored the effect of imperfections. We believe that the proposed structure may have disruptive applications as an integrated low-power, low-noise nanophotonic switch or flip-flop for quantum optics applications.

**Acknowledgement**

This work was supported by the Office of Naval Research, the Air Force Office of Scientific Research and the Simons Foundation. E.V. was supported by the Netherlands Organisation for Scientific Research (NWO).

**Figures**

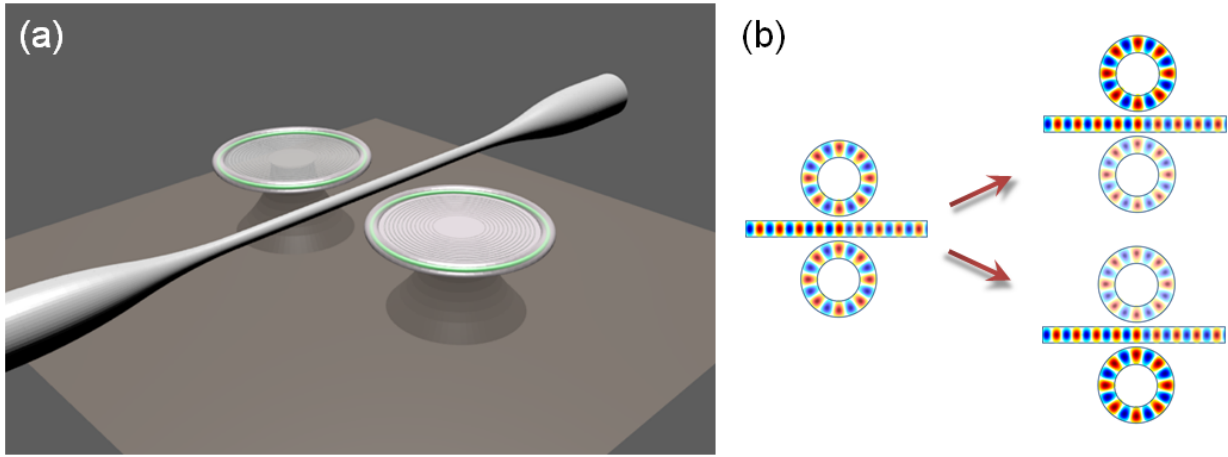

Fig. 1. (a) A symmetric arrangement of coupled optomechanical cavities, (b) Schematic representation of bifurcation and mirror symmetry breaking.

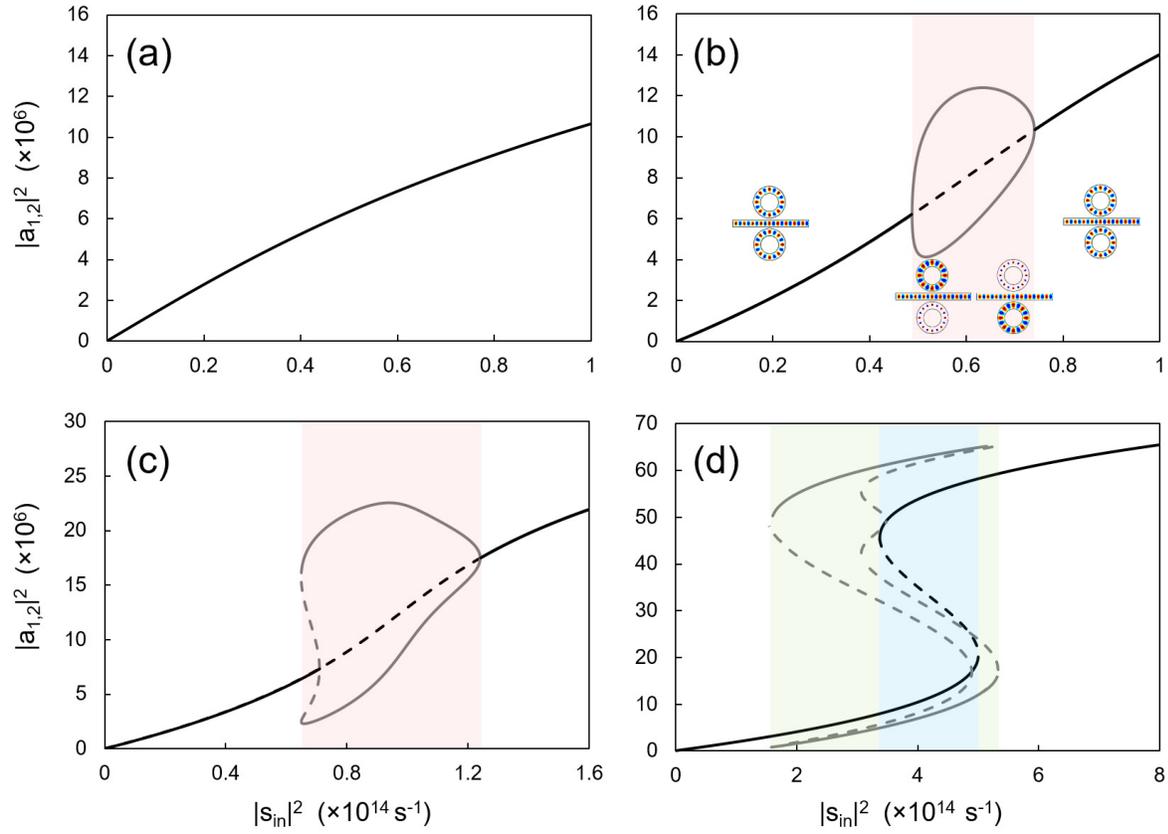

Fig. 2. The nonlinear eigenstates $|\bar{a}_{1,2}|$ of the two cavities for (a) $\Delta = 0$, (b) $\Delta = -\kappa_\ell$, (c) $\Delta = -1.5\kappa_\ell$, and (d) $\Delta = -4\kappa_\ell$ as a function of the input photon flux $|s_{in}|^2$. (b) The phase space evolution of the nonlinear eigenstates for the case of $\Delta = -\kappa_\ell$. In all cases, black and gray curves depict the symmetric and asymmetric solutions respectively while the solid and dashed curves, on the other hand, represent the stable and unstable regions. Light red, green and blue regions respectively represents regions with two, three and four stable eigenstates. The parameters used for these simulations are $\kappa/2\pi = 2\kappa_e/2\pi = 2\kappa_\ell/2\pi = 1$ MHz, $\Omega_m/2\pi = 50$ MHz, $\Gamma_m/2\pi = 10$ KHz, $G/2\pi = 6$ GHz/nm, and $m = 6$ ng.

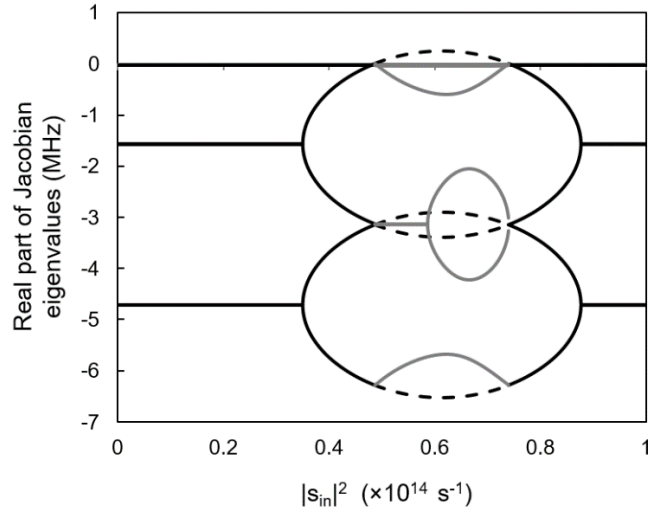

Fig. 3. (a) Real part of the Jacobian eigenvalues to investigate the stability of the nonlinear eigenstates shown in Fig. 2(b) for $\Delta = -\kappa_\ell$. Here the solid black and grey lines represent the symmetric and asymmetric regions respectively, while the dashed line is associated with the symmetric branch in the region where it coexist with the asymmetric solution. The only portion with positive values corresponds to symmetric eigenstates in the power range where symmetry breaking occurs.

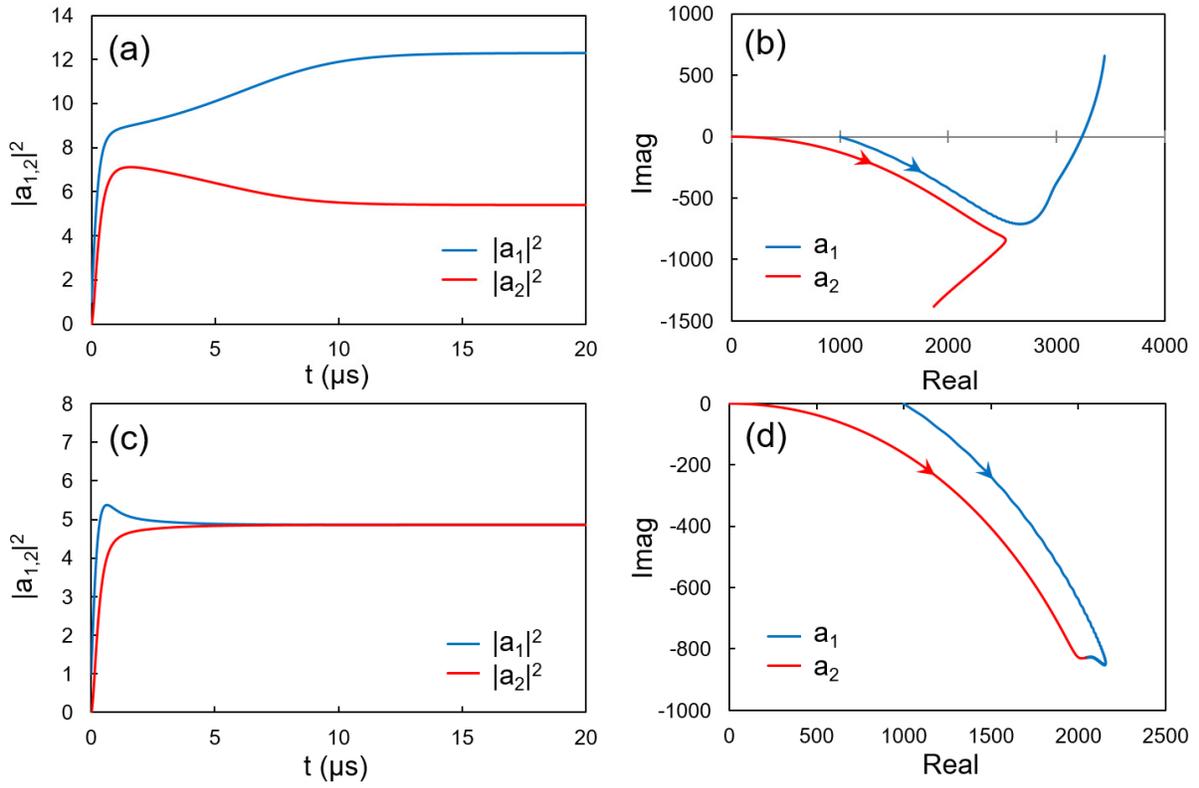

Fig. 4. Temporal dynamics of the intra-cavity photon numbers $|a_{1,2}|^2$ and the evolution of $a_{1,2}$ in the phase space for broken (a,b) and unbroken (c,d) symmetry regimes. In all cases, blue and red colors correspond to the first and second cavity respectively. Here, $\Delta = -\kappa_\ell$ and for (a,b) $|s_{in}|^2 = 0.6 \times 10^{14}\ s^{-1}$ while for (c,d) $|s_{in}|^2 = 0.4 \times 10^{14}\ s^{-1}$.

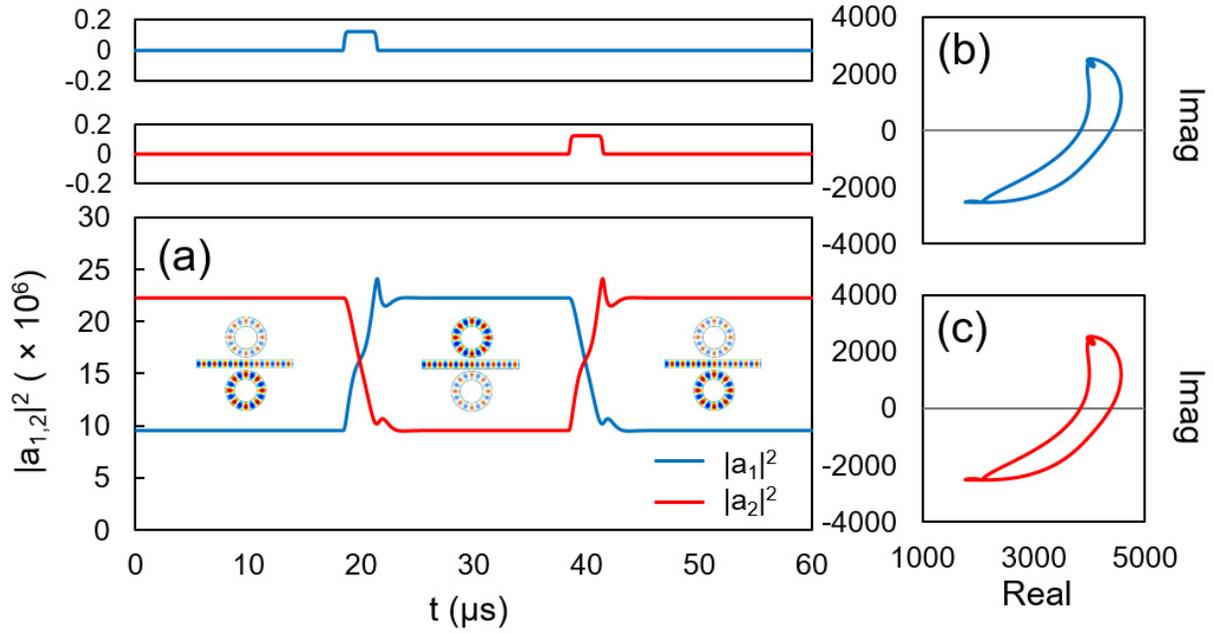

Fig. 5. (a) Time domain dynamics of the normalized optical mode amplitudes. Being in one of its two stable steady states, the system can switch to the other state by injecting small pulses to the cavities. The top panels depict trigger pulses built up in the two cavities which could be excited from a separate channel. (b,c) Phase space evolution of $a_1$ and $a_2$. The parameters used for these simulations are the same as Fig.2(c) ($\Delta = -1.5\kappa_\ell$) while the input photon flux is assumed to be $|s_{\text{in}}|^2 = 10^{14}$ s$^{-1}$.

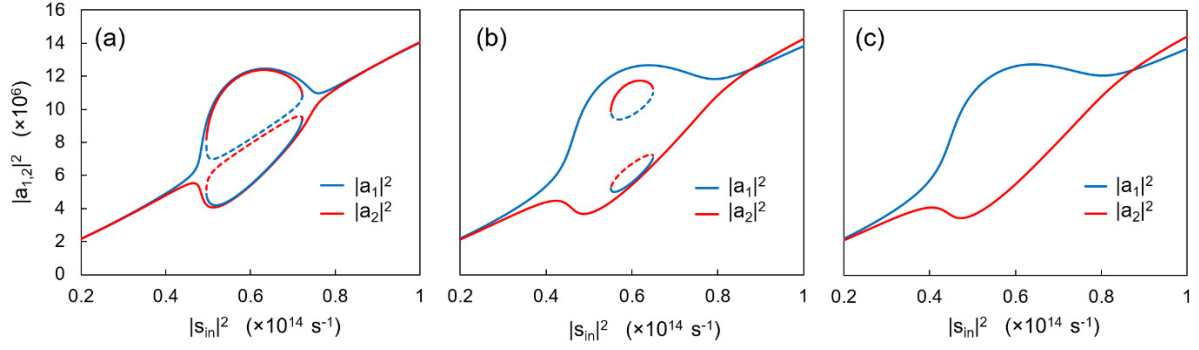

Fig. 6. The nonlinear eigenstates when perturbing the optomechanical nonlinearities of the two cavities to $\gamma_1 = (1+\varepsilon)\gamma$ and $\gamma_2 = (1-\varepsilon)\gamma$, where, (a) $\varepsilon = 0.002$, (b) $\varepsilon = 0.015$ and (c) $\varepsilon = 0.025$. In all cases, blue and red curves correspond to the first and second cavities respectively, while the solid and dashed lines represent stable and unstable modes. All parameters are the same as in Fig. 2(b).